# The Effects of Trade Openness on $CO_2$ Emission in Vietnam

Le Thi Thanh Mai[1], Le Hoang Anh[1], Kim Taegi[*1]

[1] *Department of Economics, College of Business Administration, Chonnam National University, Gwangju, South of Korea*

[*]*Corresponding author*

Abstract

This paper investigates the relationship between trade openness and $CO_2$ emissions in Vietnam using the data from 1986 to 2014. We examine the consistency of the environmental Kuznets curve hypothesis (EKC) and the pollution heaven hypothesis (PHH) in Vietnam case. In 1986 Vietnam government began to launch free-market economic reforms. Since then, Vietnam economy experienced the breakthrough innovation in trade openness. On the other hand, Vietnam witness a growing level of $CO_2$ emission. The annual growth rate of $CO_2$ emission during the period is 7.26%, and that of trade volume is 16.11%. The empirical results show that the relationship between $CO_2$ emissions and income per capita is an inverted U-shaped, consistent with to EKC hypothesis. We also find that the pollution heaven hypothesis is supported in that energy use and international trade contribute to air pollution, but becoming a full member of WTO brings positive effect to Vietnamese environment.

1. Introduction

Pollution heaven hypothesis (PHH) argues that developed countries tend to transfer pollution-intensive industries to developing countries. Weak environmental regulations and low industrial waste treatment cost in poor countries are factors which attract rich countries to invest in dirty manufacturing industries. Therefore, it can be suggested that trade openness and FDI inflow may be one of factors which has a negative impact on pollution in developing countries.

Environmental Kuznets Curve (EKC) was first introduced by Simon Kuznets in the 1950s and 1960s. EKC hypothesis indicates the inverted U shaped curve relationship between income and environmental pollution. It can be interpreted that as income level increase, pollution level will first rise until reach the turning point and then decline.

Many papers evaluate the impact of trade openness on Vietnamese economy. To the best of our knowledge, however, there has not been any research regarding the influence of trade openness on environment in Vietnam. On the one hand, in 1986 Vietnam government began to relax restrictions and launch free-market economic reforms with aim at attracting foreign investment, opening the goods market, trading with foreign countries. Since then, Vietnam economy experienced the breakthrough innovation in trade openness. On the other hand, recently Vietnam witness a growing level of $CO_2$ emission. We are motivated to investigate whether there is relationship between trade openness and air pollution in Vietnam through testing the validity of EKC and PHH. We attempt to shed further light on the correlation between the pollution and economic development focus on the trade openness process in Vietnam, a low average income country. This is the first paper consider the impact of trade openness events on the $CO_2$ emission in Vietnam. We use time series data of the period of 1986-2014. Representative pollutant is $CO_2$ emission.

With the purpose of testing validity of EKC hypothesis, we employ OLS regression to define the relationship between $CO_2$ emission and GDP. The outcome shows evidence to support this hypothesis, presenting the inverted U shape of this nexus. We also observe the positive correlations between international trade volume and air pollution in Vietnam, so we cannot deny the existence of PHH in Vietnam. In addition, we find that whereas joining WTO helps improving the air quality, developing the foreign trade activities contributes to the $CO_2$ emission in Vietnam. Besides, energy consumption is illustrated to cause a harmfulness to environment. From these findings, implications for Vietnamese government are setting the stricter regulations on production and using energy, restricting the exports of these pollution intensive products and encouraging the exports of green products.

The rest of paper proceeds as follows. Section 2 reviews the main literature investigating the relationship between $CO_2$ emission and economic development and trade openness. Section 3 represents an overview of trade openness process in Vietnam. In section 4, we describe the data, introduce the model and present the empirical results. The last section contains a brief conclusion.

2. Literature Review

Based on PHH and EKC theory, a number of empirical studies examine the relationship between FDI and environmental pollution. Many authors find empirical evidences that support for PHH in developing countries. Hoffmann, et al. (2005) examine the relationship between FDI and pollution on 112 countries over 15-28 years. They use Granger Causality Test to analyze panel data and conclude that FDI has positive relationship with $CO_2$ emissions in low and middle income countries, whereas the result of high income countries is not obtained. Ren, et al. (2014) apply in China case, conduct analysis on international trade, FDI and $CO_2$ emissions nexus in China's industrial sectors. By two-step GMM estimator, authors illustrate the positive relationship of FDI-$CO_2$ emissions and trade surplus-$CO_2$ emissions. Empirical results also claim the inverted U shape EKC which is represented by coefficient square of industrial sector's income per capita. Contributing to literature, Baek and Koo (2009) examine the FDI-economic growth-environment nexus in the case of China and India by employing cointegration analysis and a vector error correction model. China shows that FDI increases the $CO_2$ emissions in both long run and short run, while India states a small impact on pollution. For GDP, authors indicate that GDP makes pollution worsen. Observing the same finding, Solarin, et al. (2017) support PHH in Ghana. They use ARDL method to analyze the different time series models in Ghana for the period of 1980-2012. GDP, GDP square, energy consumption, renewable energy consumption, fossil fuel energy consumption, FDI, institutional quality, financial development, urbanization and trade openness are used as determinants of the model. Authors show that FDI, GDP, urban population, financial development and international trade increase $CO_2$ level, whereas institutional quality has negative impact on $CO_2$ emission.

Besides studies showing outcomes supporting for PHH, many authors find evidences to reject this hypothesis. Birdsall and Wheeler (1993), a study in Latin America, claims that increase foreign direct investment can make the developing countries apply cleaner industry by importing the pollution standard of developed countries. Zhu, et al. (2016) test the hypothesis in the case of five ASEAN countries (Indonesia, Malaysia, the Philippines, Singapore and Thailand) by panel quantile regression model. FDI, trade openness, industrial output and energy consumption are considered in the model. In contrast to Ren, et al. (2014), this study observes a negative relationship between FDI and environmental pollution. Trade openness also reduces the $CO_2$ emission, whereas energy consumption increases it. Another study against PHH is from Al-mulali and Tang (2013), which employ multivariate framework, and Fully Modified OLS to analyze the data of Gulf Cooperation Council countries from 1980 to 2009. The outcome is that energy consumption, trade openness, urbanization and GDP growth increase $CO_2$ emission whereas FDI inwards have a negative nexus with $CO_2$ emission in long run. Besides, using short run Granger causality test results, they find that FDI has no short run causal relationship with $CO_2$ emission.

Tamazian and Rao (2010) conduct the estimation on 24 transitional economies from 1993 to 2004. They employ the standard reduced-form modeling approach and GMM estimation for their data, then get the results supporting for the EKC hypothesis. Besides, they claim the negative effect of financial liberalization on environment if institutional framework is not controlled strictly. Another support for this hypothesis comes from Pao and Tsai (2011). The data is estimated by panel cointegration technique. In long-run equilibrium, $CO_2$ emissions appear to be energy consumption elastic and FDI inelastic, and the results seem to support the EKC hypothesis. Moreover, this study also supports for PHH.

Chandran and Tang (2013) study the relationship between energy consumption, FDI and $CO_2$ emissions for five ASEAN countries from 1971 to 2008. Using the multivariate cointegration test and Granger causality analysis, their empirical results indicate that $CO_2$ emissions and their determinants are co-integrated only in Indonesia, Malaysia and Thailand. FDI is not significant, economic growth is an important factor affecting pollution. And they conclude that their study does not support for inverted U-shape EKC hypothesis. Another study provides evidence against EKC hypothesis is from Narayan and Narayan (2010). They test the EKC hypothesis by panel data of 43 developing countries. Thirty-five percent of samples indicate falling pollution over the long run. Besides, testing in Middle Eastern and South Asian show the negative nexus among income and $CO_2$ emissions.

3. An overview of trade openness (TO) in Vietnam

The war in Vietnam officially ended in 1975, and its consequence was heavy, the economy was ruined. Right after war, because of political mechanism, Vietnam was isolated from the world. Pursuing socialist economy, Vietnamese government closed the economy, banned private businesses and built state trading network.

Passing 1970s, realizing that the economy policy was ineffective, the politicians began opening the economy. "Doimoi" policy was launched with aim at transition from a centralized economy to an open market economy. In that period, private business was encouraged, regulations for foreign investor was loosen, and economic relations were opened. Since then, Vietnamese economy gradually integrated with the world. Vietnam became a member of Association of Southeast Asian Nations (ASEAN) in 1995 and Asia-Pacific Economic Cooperation (APEC) in 1998. As a member of ASEAN, Vietnam has participated in the recently established FTAs between ASEAN and Japan, China and Korea.

In 2001, U.S.-Vietnam Bilateral Trading Agreement (US- Vietnam BTA) was signed, marked the end of cold war period between Vietnam and USA. The US-Vietnam BTA placed an important role in Vietnamese economic integration by "spurring political will to speed up negotiations on Vietnam's accession to WTO" in later years (CIEM-USAID, 2007). In 2007, Vietnam has been officially a full member of World Trade Organization (WTO). As the regulations of WTO, Vietnam government has to open the market, cuts down import tariff, removes tariff barriers overtime as schedule committed with WTO. After joining to WTO, Vietnamese customers have chances to consume the large amount of import goods, and enterprises can introduce their products into the world market. Besides, FDI inflow was also increasing significantly and reached an all-time highest of US$71.7 billion in 2008.

In our research, the year 2007 when Vietnam join to WTO is chosen as a breakthrough event in this process. Additionally, we also employed trade volume of import and export as a measure of trade openness.

4. The model and Estimation

4.1. Data

This study covers the period from 1986 to 2014 and the data is extracted from World Bank database. A descriptive analysis of the series is conducted in Table 1, which presents observation numbers, mean, standard deviation, coefficient of variation and exponential growth rate in this work.

All variables are collected yearly from 1986 to 2014, that leads each variable has 29 observations. Carbon dioxide emission remained stable at the low points in late 1980s and early 1990s and then was increasing since 1996 and reached the peak in 2014. In general, the growth rate of $CO_2$ emission is 4.02% annually and average level is 872.1 tons per capita, relatively low in the region. Regarding to income, Vietnam experienced optimistic development in this period, has transferred from poorest countries group to lower middle-income nation. The average annual GDP per capita is at $688.1 and increases 8.59% per year. GDP varies widely among the years, that is indicated by coefficient of variation 82.01%.

**Table1**: Descriptive analysis in level form of variables

| Variable | Observation | Mean | Standard Dev. | Coefficient of variation(%) | Growth rate (%) |
|---|---|---|---|---|---|
| $CO_2$ | 29 | 872.1 | 521.5 | 59.80 | 7.26 |
| GDP | 29 | 688.1 | 564.3 | 82.01 | 8.59 |
| FDI | 29 | 37.1 | 37.2 | 100.19 | 24.59 |
| Trade | 29 | 76.8 | 91,4 | 119.06 | 16.11 |
| EU | 28 | 406.5 | 144.9 | 35.66 | 4.02 |

Note: The unit of CO2 is metric tons per capita. The units of GDP and FDI are current US$ per capita. The unit of Trade is current billions US$. The unit of EU is kg of oil equivalent per capita.

Since "Doimoi" policy was launched in 1986 with aim at opening the market, Vietnam began attracting inflow FDI with the starting point of $40,000 in 1986. The period from 1990 to 1996, more FDI is invested notably into Vietnam market and contribute significantly to Vietnamese economic development. It is demonstrated by high value of coefficient of variation at 100.19% as

well as growth rate at 24.59%. Although the slight decline as a result of Asian financial crisis 1997 and competition from other countries in region especially China, Vietnam observed the enormous FDI inflow since 2007 when Vietnam officially jointed the WTO, reached the peak of over $9.5 billion FDI. Thanks to "Doimoi" policy, along with FDI, openness policy helps trade volume in Vietnam rise rapidly among the years with coefficient of variation 119.06%. Over 29 years, import and export volume achieved annual growth rate at 16.11%. From the smallest trade volume in 1988 at 18.95% of GDP, Vietnam achieved the top at 169.53% of GDP in 2014 in this period. Concerning energy use, Vietnam consumes average 406.5 kg of oil equivalent per capita annually through the considered period. Although the power consumption increases over time as general trend, the growth speed experienced reasonable rate at 4.02%.

### 4.2. Model specifications

Environmental Kuznets Curve (EKC) was first introduced by Simon Kuznets in the 1950s and 1960s, and then Grossman and Krueger (1995) tested this theory by the model of relationship between pollution level and income and square of income. The model (1) is conducted to test the validity of EKC in Vietnam case.

$$lnCO2_t = \beta_0 + \beta_1 lnGDP_t + \beta_2 lnGDP_t^2 + \varepsilon_t \tag{1}$$

Where $CO2_t$ is carbon dioxide emissions (metric tons) per capita, $GDP_t$ represents gross domestic product per capita (current US$) and $GDP_t^2$ is the square of per capita gross domestic product. We use the variable $GDP_t^2$ with aim at testing the validity of Environmental Kuznets Curve hypothesis for all models. To support EKC, $\beta_1$ is expected to be positive whereas $\beta_2$ should be negative. At that time, the relationship between income and pollution will be presented as inverted U shape curve. Whereas, the following model aim at examining the Pollution Haven Hypothesis in Vietnam.

$$lnCO2_t = \beta_0 + \beta_1 lnGDP_t + \beta_2 lnGDP_t^2 + \beta_3 lnEU_t + \beta_4 lnFDI_t + \beta_5 lnTrade_t + \beta_6 TO_t + \beta_7 T + \varepsilon_t$$

(2)

Trade liberalization opens a great opportunity to import and export between countries through Free Trade Agreements (FTAs), as well as a significant inflow FDI into Vietnam. The event joining WTO in 2007 is a breakthrough of trade openness issue in this country. In the equation (2), FDI (current US$ per capita) is represented by $FDI_t$, while $Trade_t$ is total volume of import and export, and $TO_t$ is dummy variable representing the trade openness which is given 0 for the year before 2007 and 1 for the year from 2007. To be consistent with PHH hypothesis, we predict the coefficients of $FDI$, $Trade$ and $TO$ are positive, indicating that $CO_2$ emission increases as FDI inflow, trade volume increase and the trade openness has a negative impact on Vietnam environment.

The literature on the environment matter has proved that energy consumption causes a negative impact on environment. Al-mulali and Tang (2013) considered energy consumption as a determinent in their model. The process of producing power as well as consuming energy release a number of carbon dioxide to the air. We therefore, consider energy use $EU_t$ (kg of oil equivalent per capita) as an element causing the increase of $CO_2$ emission.

All models are estimated by ordinary least square (OLS) regression with logarithm form of $CO_2$, $GDP$, $GDP^2$, $FDI$, $EU$, and $Trade$.

### 4.3. Estimation Results

Table 2 provides the regression results of models testing the validity of EKC hypothesis and PHH in Vietnam case from 1986 to 2014. Coefficients, R-squared and adjust R-squared are represented in the table, t-values are in parentheses.

Table 2 indicates that EKC hypothesis is valid in the Vietnam case that illustrated by significant positive and negative coefficients of $GDP$ and $GDP^2$ respectively. This also consistent with previous studies of Pao and Tsai (2011), Tamazian and Rao (2010), Tang and Tan (2015). Figure 1 shows that the GDP in Vietnam is currently around the peak (approximately 2000$U per capita) of the EKC and $CO_2$ emissions likely move down with an increase in GDP per capita. Further, model (1) shows that the air quality of Vietnam turns to worse over time by significant positive coefficients of time trend variable at 1% level.

**Figure 1**: Environmental Kuznets Curve: Case in Vietnam

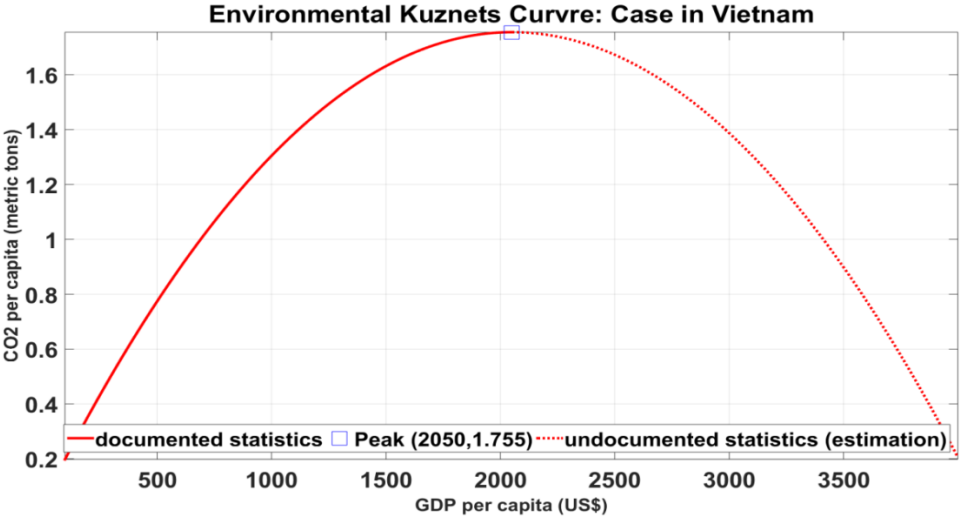

Note: The graph is drawn by using the regression results of equation (1).

Adding energy use variable makes the explanation power of model stronger, with higher value of $R^2$ and adjusted-$R^2$. Table 2 provides the evidence that energy consumption causes the air pollution in Vietnam. All coefficients from model (2) to model (6) are positive and significant at 1% level. The harmfulness of using energy is obvious and proved in previous studies, when fossil and fuels energy is current main power used in the world.

With regard to FDI, there is no evidence that FDI has impact on $CO_2$ emission in our outcomes, represented by insignificant negative coefficients. This result is similar with the work of Tang and

Tan (2015) which conducted the hypothesis using data in Vietnam from 1976 to 2009. Their finding yields a negative impact of FDI on $CO_2$ emission but it is not significant statistically. An explanation for this finding might be that Vietnam's economy is a closed economy before "Doimoi" since 1986, FDI is likely negative from 1976 to 1985 and increased inconsiderably until 1990. Therefore, it is hard to conclude that the relationship between $CO_2$ emission and FDI in this period is significant. FDI inflows in Vietnam just increases rapidly since 2006 when the market is more liberalization and, in particular, Vietnam joined the WTO.

**Table 2**: Regression result

| Variable | Model (1) | Model (2) | Model (3) | Model (4) | Model (5) | Model (6) |
|---|---|---|---|---|---|---|
| Constant | 2.4220** | -8.7601*** | -7.4654*** | -12.7174*** | -14.0621*** | -13.6460*** |
|  | (2.50) | (-4.21) | (-3.20) | (-3.73) | (-4.83) | (-4.51) |
| $lnGDP$ | 0.8548** | 1.5628*** | 1.5387*** | 1.3531*** | 0.7458** | 0.7803** |
|  | (2.65) | (5.92) | (5.86) | (5.15) | (2.42) | (2.46) |
| $lnGDP^2$ | -0.0490* | -0.1293*** | -0.1273*** | -0.1219*** | -0.0683** | -0.0718** |
|  | (-1.80) | (-5.29) | (-5.24) | (-5.32) | (-2.55) | (-2.59) |
| $lnEU$ |  | 1.7812*** | 1.5500*** | 1.6817*** | 2.0340*** | 1.9212*** |
|  |  | (5.81) | (4.29) | (4.87) | (7.78) | (6.06) |
| $lnFDI$ |  |  | -0.0146 | -0.0182 |  | -0.0072 |
|  |  |  | (-1.18) | (-1.55) |  | (-0.65) |
| $lnTrade$ |  |  |  | 0.2440* | 0.2968** | 0.3012** |
|  |  |  |  | (2.01) | (2.77) | (2.76) |
| $TO$ |  |  |  |  | -0.2013*** | -0.1846** |
|  |  |  |  |  | (-3.16) | (-2.65) |
| $R^2$ | 0.9737 | 0.9888 | 0.9894 | 0.9911 | 0.9933 | 0.9934 |
| Adj. $R^2$ | 0.9706 | 0.9868 | 0.9870 | 0.9886 | 0.9914 | 0.9911 |

Note: 1) (*), (**) and (***) indicate significance at the 10%, 5% and 1% level respectively. 2) The regression results in the Table are estimated with including time trend variable T (not reported).

It is evident from the table 2 that trade openness, represented by total volume of international trade, contributes to $CO_2$ release in Vietnam, which are observed from estimation results of model (5) and model (6) at 5% significant level, giving evidences to support PHH. The increasing trade between Vietnam and foreign countries leads the growth of $CO_2$ release. Increased trade means expanded production and consumption, which contribute to the air pollution from the waste of these activities. Observing the same finding, Ren, et al. (2014), Solarin, et al. (2017) also recognize these relationships in China's industrial sectors and in Ghana respectively. However, this result is in sharp contrast with finding of Zhu, et al. (2016). They find that trade openness has a negative impact on carbon emissions, indicating that a higher level of trade openness can relieve carbon emissions in low- or high-emissions countries.

Further, for testing the impact of trade openness on environment in Vietnam, we also examine the dummy variable TO which equals to 0 before the event Vietnam become a full member of WTO and equals to 1 after the event. Table 2 exhibits the positive impact on $CO_2$ emission of trade openness at 5% significant level. This proves that air quality of Vietnam becomes better after joining the WTO. The explanation is that to become a full member of WTO, Vietnam has to construct the better regulations system which contributes to environment protection. The value of coefficients of dummy variable, however, are lower than trade volume, therefore in overall trade openness causes a slightly pollution on Vietnamese environment.

5. Conclusions

The impact of economic development and environment pollution attracts interest of many scholars. The environmental Kuznets curve hypothesis (EKC) and the pollution heaven hypothesis (PHH) are conducted in different countries with various methodologies. However, there are still debates on the inverted U-shape of the EKC and the PHH is rejected by some empirical evidences. Our paper tests these hypotheses in Vietnam with an additional consider dummy variable representing trade openness in Vietnam. We find that the relationship between GDP and $CO_2$ emissions in Vietnam is consistent with the EKC hypothesis, and international trade has positive correlation with $CO_2$ emissions which also support for the PHH. There are two interpretations of this outcome. First, weak environmental policy is considered as a source of comparative advantage for low income countries to attract foreign investors. Second, weak management ability of regulators and limited monitoring cost may facilitate for high income countries to shift pollution-intensive industries to poorer countries. However, joining WTO helps Vietnam reduce the $CO_2$ level, because the regulations are more completed. On the other hand, energy use is proved to be a determinant causing air pollution.

A2. Variable explanation

| Variables | Definition | Source | Calculation | Unit | Expectation |
|---|---|---|---|---|---|
| $CO_2$ | $CO_2$ emissions | WDI | | Metric tons per capita | |
| $GDP$ | Gross Domestic Production per Capita | WDI | | Current US$ per capita | Positive impact on CO2 emission |
| $FDI$ | Foreign direct investment per Capita, net inflows | WDI | | Current US$ per capita | Positive impact on CO2 emission |
| $EU$ | Energy use | WDI | | Kg of oil equivalent per capita | Positive impact on CO2 emission |
| $Trade$ | Total volume of import and export | WDI | Import+Export | Current US$ | Positive impact on CO2 emission |
| Import | Imports of goods and services | WDI | | Current US$ | |
| Export | Exports of goods and services | WDI | | Current US$ | |
| $TO$ | Trade openness | WDI | $TO$=0 for the year before 2007<br>$TO$=1 for the year from 2007 | | Positive impact on CO2 emission |